# LOCALIZED LNA COOLING IN VACUUM

*Frans Schreuder, Jan Geralt Bij de Vaate*
ASTRON, P.O. Box 2, 7990 AA Dwingeloo, The Netherlands.
schreuder@astron.nl

## ABSTRACT

In the Square Kilometre Array (SKA) telescope [1], [2], the noise temperature of the first LNA must be reduced in order to reduce the necessary active area and the total system costs.
Cooling the LNA locally would significantly decrease the noise figure but also the necessary power since not the whole system has to be cooled.
For optimal thermal isolation, an LNA chip which only needs 6 bondwires has been chosen, 4 Ground and 2 signal wires. Biasing occurs on-chip. If the bondwires are 1.5mm long, the total heat conduction of the 6 bondwires is 31 mW, which is added to the power consumption of the LNA (30 mW). With a power of 61 mW to cool, the Peltier element can achieve a ΔT of 60K.
With this system, a noise reduction of 30% has been measured with 0.5W of electrical power. For 15% noise reduction, only 35mW of electrical power was needed.

## 1. INTRODUCTION

For the Square Kilometre Array telescope (SKA), over 100 million antenna elements will be combined into one big telescope. The sensitivity of the telescope increases with the active area (Aeff), but it decreases with the noise temperature of the system (Tsys). The parameter Aeff /Tsys Is defined to be 20000; with a 50K system noise temperature (≈0.7dB Noise Figure) the total effective area of the telescope will be about 1 square kilometre.
If the noise temperature of the first LNA will be reduced, the effective area can also be reduced which results in lower costs for the SKA telescope.
To give an LNA (Low Noise Amplifier) a better noise figure, it is possible to cool down the chip.
In earlier investigations, a cooling method has been investigated where Peltier elements were used. The results of these investigations were insufficient because of the thermal leak to the PCB.
The title "Localized LNA cooling in vacuum" means that not the whole system is cooled but just the LNA chip in which most of the noise is generated.
The main question of investigation is: "Is localized LNA cooling in vacuum a realistic option to implement into SKA?"

## 2. TEMPERATURE DEPENDENCY OF THE NOISE FIGURE

For a perfect resistor, the noise depends linearly on the temperature since $P_n$=*kTB*. But a transistor is not purely resistive.
An often used noise model for HEMT transistors is Pospieszalski's two-temperature model where the noise sources in the transistor are two resistors $R_g$ and $R_{ds}$ with a temperature response $P_n$=*kTB* [4], [5].
The two noise sources are correlated and cause a total noise shown in equation 1. The temperatures $T_g$ and $T_d$ are not real temperatures, but equivalent noise temperatures of the resistors, though they are both proportional to the junction temperature.

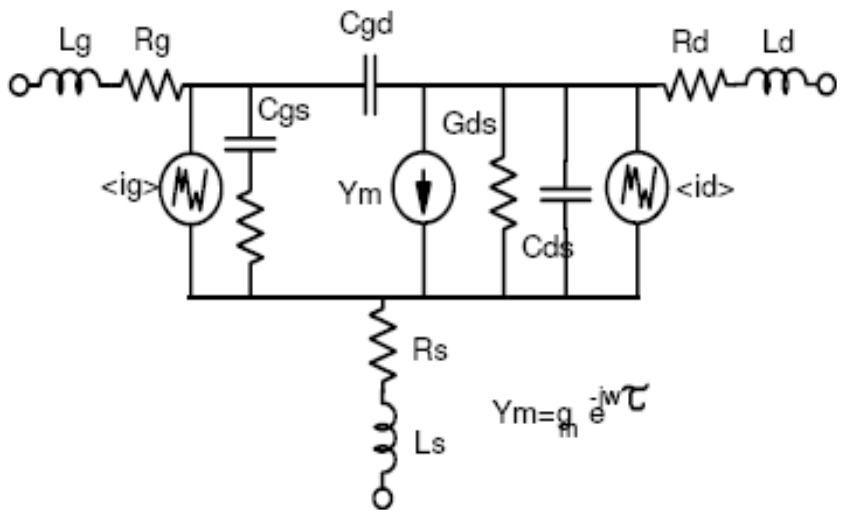

**Figure 1: Small signal and noise equivalence circuit.**

$$T_n = 2 \cdot B \cdot \frac{f}{f_T} \cdot \sqrt{r_t \cdot T_g \cdot g_{ds} \cdot T_d} \qquad 1$$

Equation 1 contains a $\sqrt{T^2}$ , so the relation of $T_n$ versus $T_{junction}$ is purely proportional for thermal noise.
In LNA5, an MMIC LNA designed by Roel Witvers which is used in all the measurements, the constants $g_{ds}$ and $r_t$ are unknown, but what is known is the noise temperature at $T_{junction}=293K$.
Now we know $T_n$ at *293K* and that $T_n$ has a proportional relation with temperature, it is simple to derive the equation for the noise temperature as a function of $T_{junction}$ and $T_{n_293K}$.

$$T_n = \frac{T_{n_293K}}{293} \cdot T \qquad 2$$

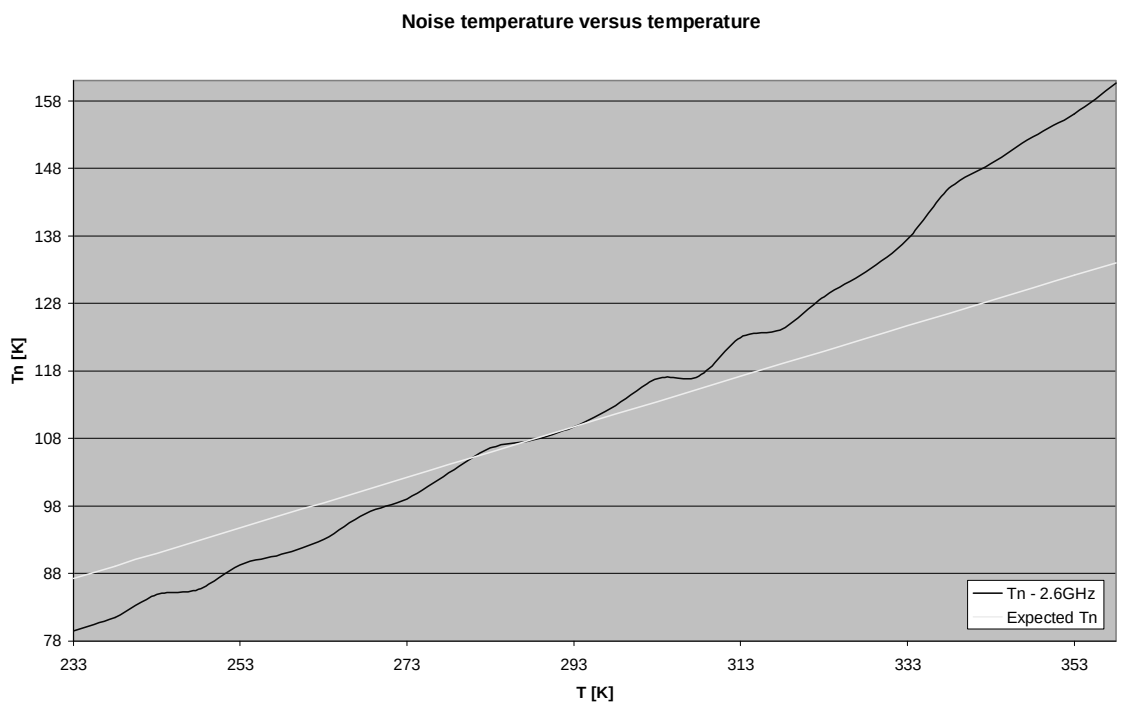

**Figure 2: Noise temperature vs. Temperature and noise temperature according to equation 2.**

The temperature of LNA 5 has been controlled in a temperature chamber over a high temperature range (-40°C to 85°C) and the noise has been measured. The graph in Figure 2 demonstrates that the noise temperature slope over the temperature is not purely proportional, but it contains a 2nd order component.

The additional slope of the noise temperature is caused by an additional gate leakage dc-current which causes shot noise. This gate-current also increases with the junction temperature. According to an investigation of Glenn Jones (Thermoelectric cooling of LNAs to 200K), this current increases significantly above 250K for an 80um InP hemt from NGST, which also has been used in LNA5 [3].

## 3. THERMAL MODELLING

Before the heat flows in the system can be defined, it is better to take a look at the mechanical drawing of the system. Figure 3 shows the different components of the thermal model, a photograph with a magnification of the chip is shown in Figure 5
The first heat source is the LNA chip, the heat that is generated in the LNA from the electrical power will flow through the TEC. Another source of heat is the conduction through the bondwires, because there will be a difference in temperature between the LNA and the heat sink. The last heat source will enter the LNA through radiation; convection through air is not possible because the system will be placed in a vacuum chamber.
The Peltier element will generate a temperature difference and the heat energy will be pumped from the cold side (LNA) to the hot side (Heat sink). Because the Peltier element also consumes electrical energy, additional heat will be generated at the hot side of the Peltier element.

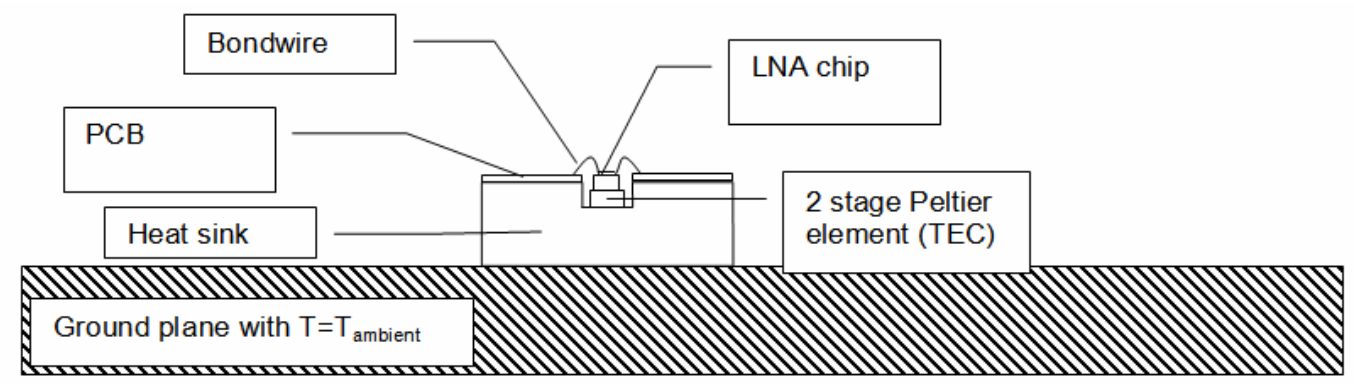

**Figure 3: Setup of the TEC cooling system**

The heat flows through the different components can be represented as electrical currents through thermal resistances. To make the heat flow more visible, an "electrical" schematic of the circuit has been made.

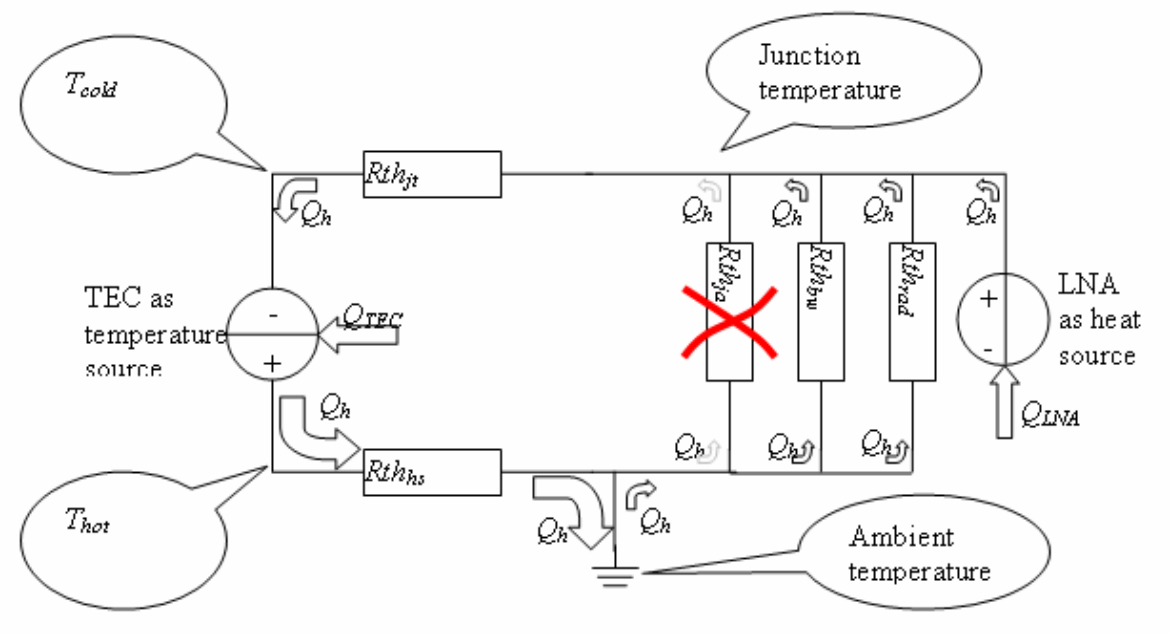

**Figure 4: Electrical representation of the heat flow**

In Figure 4, the electrical representation of the heat flow is shown, the heat ($Q_h$ *[W]*) is represented as current, the temperature (*T [K]*) is represented as voltage and the thermal resistance ($R_{th}$ *[K/W]*) is represented as electrical resistance. A major difference must be taken in mind: the active sources (TEC and LNA) can introduce additional $Q_h$ into the circuit since electrical energy is converted to heat energy at these spots. The different components in the circuit represent the following:

$Rth_{jt}$: Thermal resistance Junction – TEC
$Rth_{hs}$: Thermal resistance Heatsink
$Rth_{ja}$: Thermal resistance Junction – Ambient
$Rth_{bw}$: Thermal resistance bondwires
$Rth_{rad}$: Thermal resistance radiation
*TEC*: Thermoelectric temperature source
*LNA*: Cooled object as heat source

$Rth_{ja}$ is disabled because of the vacuum around the system, no heat can flow through convection, so the thermal resistance is infinite.

## 4. HEAT FLOW THROUGH THE TEC

In most of the cooling systems, where the temperature of the device is still above ambient, it is essential to have as much heat conductivity to ambient as possible, but as soon as the temperature comes below ambient temperature, every direct heat path (except the heat path that is going through the active cooler or Peltier element) becomes a thermal leak and will cause the device temperature to rise. In order to know which temperature can be reached with the TEC (Thermo-Electric Cooler) or even before the TEC can be ordered, the heat that will flow from the cold side to the hot side of the TEC has to be determined.
The heat in Watts that has to be pumped from the LNA to the heat sink on the bottom of the vacuum package is the sum of three heat sources:

- The heat produced in the LNA
- The heat conductivity through the bondwires
- The heat radiation through vacuum

### 4.1. Heat produced in the LNA

On top of the TEC lies LNA5, an LNA that normally operates at 3V / 35mA. For cooling purposes, the LNA should dissipate as little power as possible; this LNA has still a good performance if it is supplied with 1.5V / 20mA. The gain becomes a little lower at these low voltages, but the noise figure becomes even better and the power consumption is only 30 mW.
The heat that will be produced by the LNA is simply generated by the DC power of the supply.

### 4.2. Heat conductivity through the bondwires

The LNA, which lies on top of the TEC, could achieve the lowest temperature if it were completely isolated from the ambient, e.g. in vacuum. Unfortunately there has to be an electronic connection for the ground and the signal input and output. LNA5 does not need additional biasing wires because the DC biasing occurs through the output line. This way, only 6 bondwires are needed; one for input, one for output and both signal wires are “guided” by two ground wires, to make it a sort of coplanar waveguide.
Heat that is transferred through one bondwire of the LNA is given by the following equation:

$$Q_{cond} = \lambda(T) A \tfrac{dT}{dx} \qquad (3)$$

Equation 3 can be filled in, assuming that the bondwire is 31 mm long 25 um thick, a realistic temperature difference is 50K and there are 6 bondwires. The equation has to be multiplied by 6, because the 6 bondwires are treated as parallel (thermal) resistors. The result of this equation is 31 mW.

### 4.3. Heat radiation through vacuum

The room around the LNA will be in vacuum because it isolates the chip, and it omits ice generation.
The heat that is transferred through vacuum by radiation is given by the following equation:

$$Q_{rad} = e\sigma A T^4 \qquad (4)$$

With an area of the chip of 9 mm$^2$, the radiated power will be less than 3 mW which is negligible to other heat sources.

### 4.4. Total heat transfer through the TEC

Now the three heat sources are known, the total heat transfer can be calculated by adding the three values.

$$Q_{tot} = Q_{gen} + Q_{cond} + Q_{rad} = 30mW + 31mW + 3.0mW = 64mW \qquad (5)$$

Half of the heat power that is inserted into the cold side of the TEC is generated by the LNA, the other half comes through the bondwires. Longer or thinner bondwires would result in a better performance of the TEC but worse RF performance.

### 4.5. Vacuum vs. open air

A vacuum is for isolation purposes the best option because no convection can occur. Surrounding the chip with air in stead of a vacuum would result in a temperature rise of 5 to 10 Kelvin due to an extra heat source of approximately 27mW according to equation 6.

$$Q_{conv} = h \cdot A \cdot \Delta T \quad [W] \qquad (6)$$

$$h = 3.43 \cdot \Delta T^{1/4} \cdot A^{1/8} \quad [W \cdot m^{-2} \cdot K^{-1}]$$

Another advantage of vacuum is that moisture and ice can not be generated on the chip surface. Ice generation on the chip surface does significantly influence the electric performance and it can even destroy the chip or the bondwires.

## 5. TEST SETUP

Measurements have been done in a large vacuum chamber, this way no vacuum package was needed and the temperature could be monitored easily. Also different environment conditions from vacuum to 1 bar could be researched in this setup.

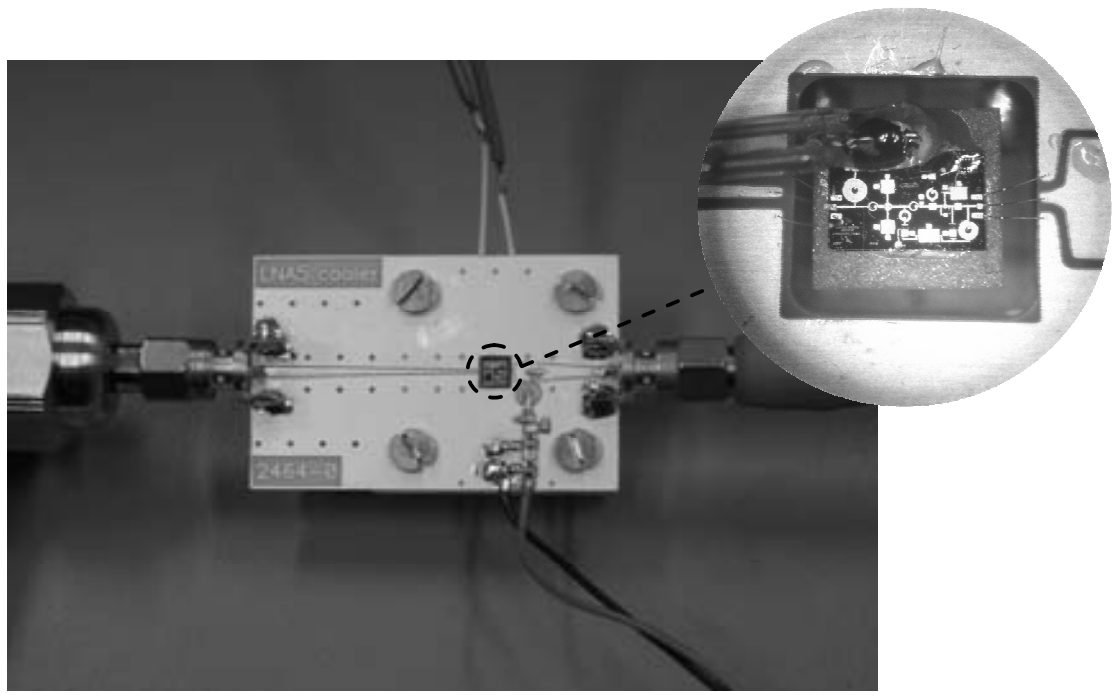

**Figure 5: Photograph of the test setup**

An aluminum cooling block has been mounted together with the Peltier, the PCB and the LNA. In this version of the system, the whole system will be placed in a vacuum chamber and no vacuum package is applied in this design stage.
The aluminum cooling block has a double function; it holds the Peltier element, so that the top of it is at the same height as the top of the PCB. This way, the LNA can be easily connected to the PCB making use of bondwires. The other function of the aluminum block is to transport the generated heat generated to ambient.
The PCB contains tapered transmission lines for impedance matching and a bias T.

## 6. IMPLEMENTATION IN THE SYSTEM

The system "localized LNA cooling in vacuum" has to be implemented into the square kilometre array. Since millions of antenna elements in this telescope will be amplified, also millions of local coolers must be applied so the price must be kept as low as possible.
For this purpose, a vacuum package has to be chosen in which standard bonding technique can be used and it has to retain a vacuum.
A standard ceramic butterfly package has been found that can contain the Thermoelectric cooler, the chip and the bondwires with proper RF impedance.

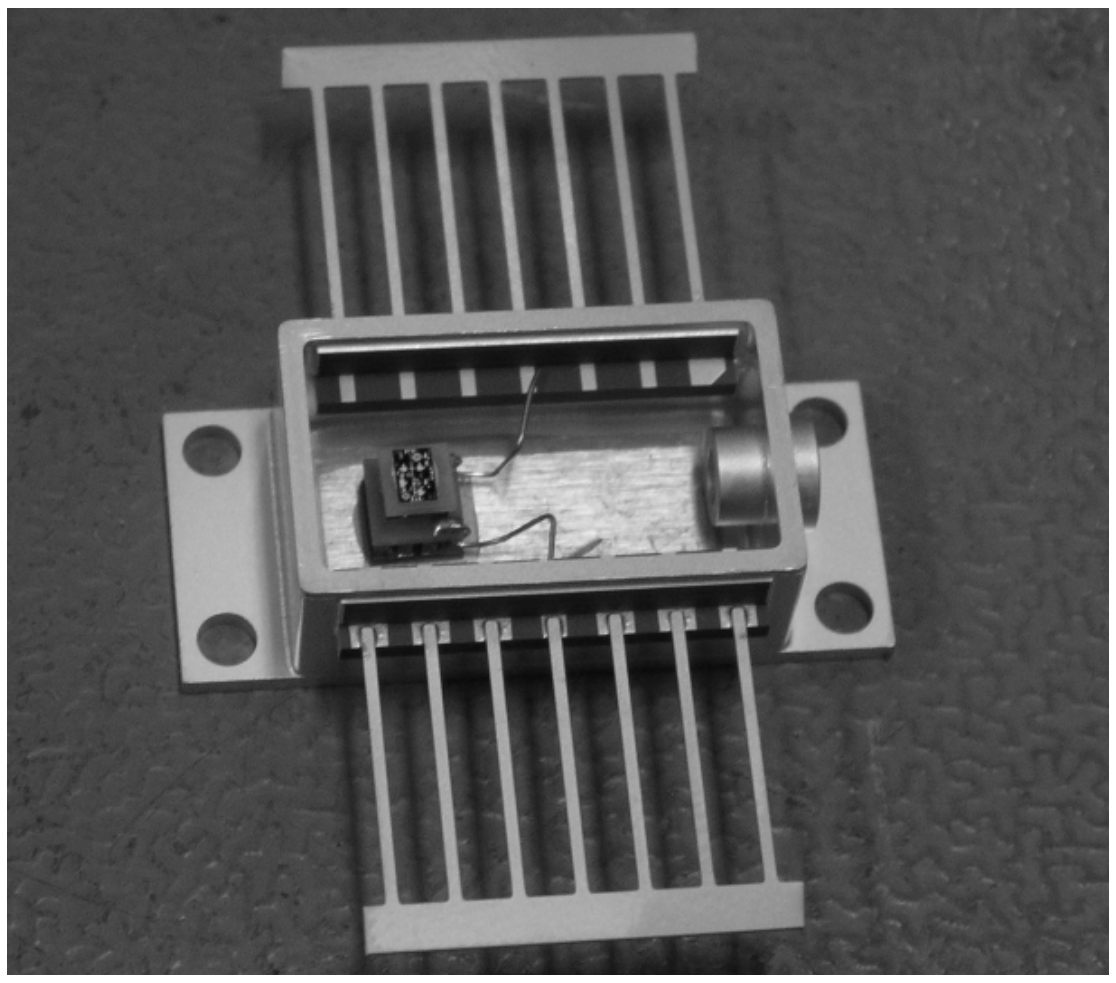

**Figure 6: TEC with LNA in a 14 pin Butterfly package**

The butterfly package is designed for a laser diode, but the hole can be used to solder a copper tube in order to close the vacuum.
The butterfly package can be mounted directly on the antenna board to avoid extra noise due to long transmission lines before the first LNA.

## 7. MEASUREMENTS RESULTS

The temperature change of the Peltier element does influence the noise figure of the LNA according to the measurements that have been applied to the system.

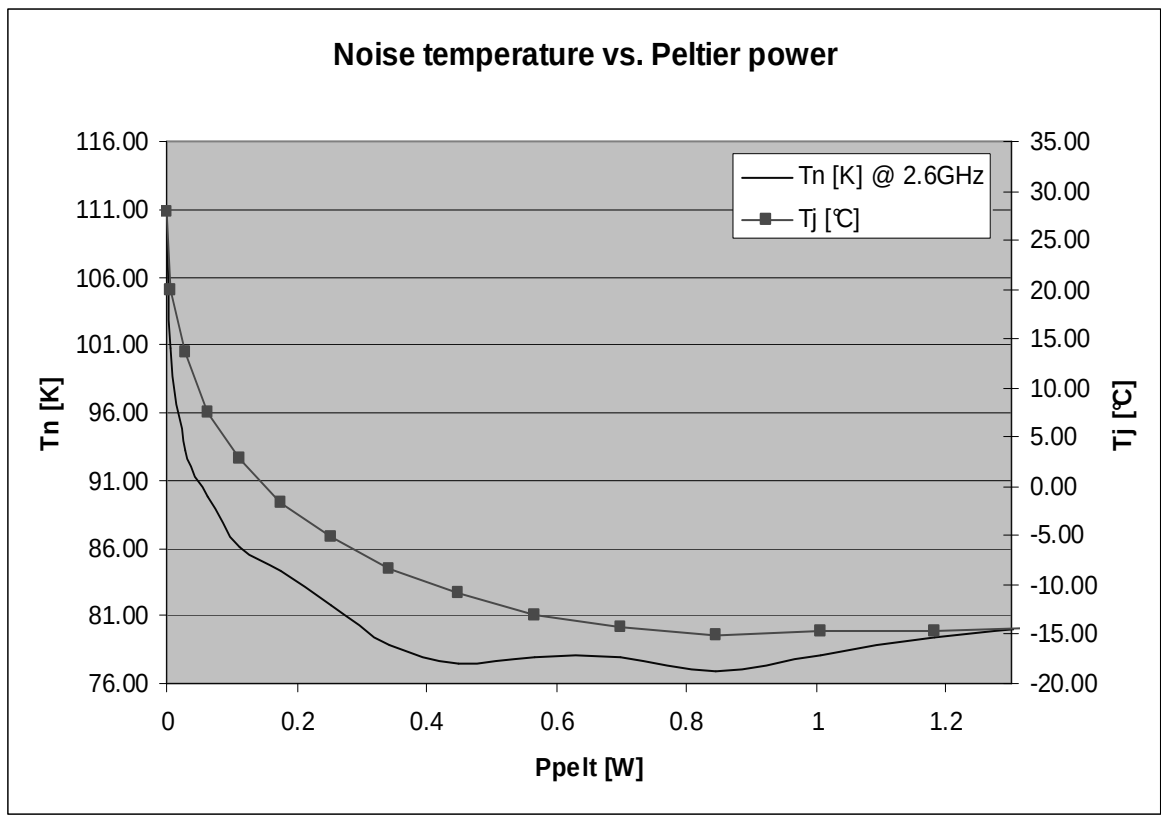

**Figure 7: Noise temperature vs. Peltier power**

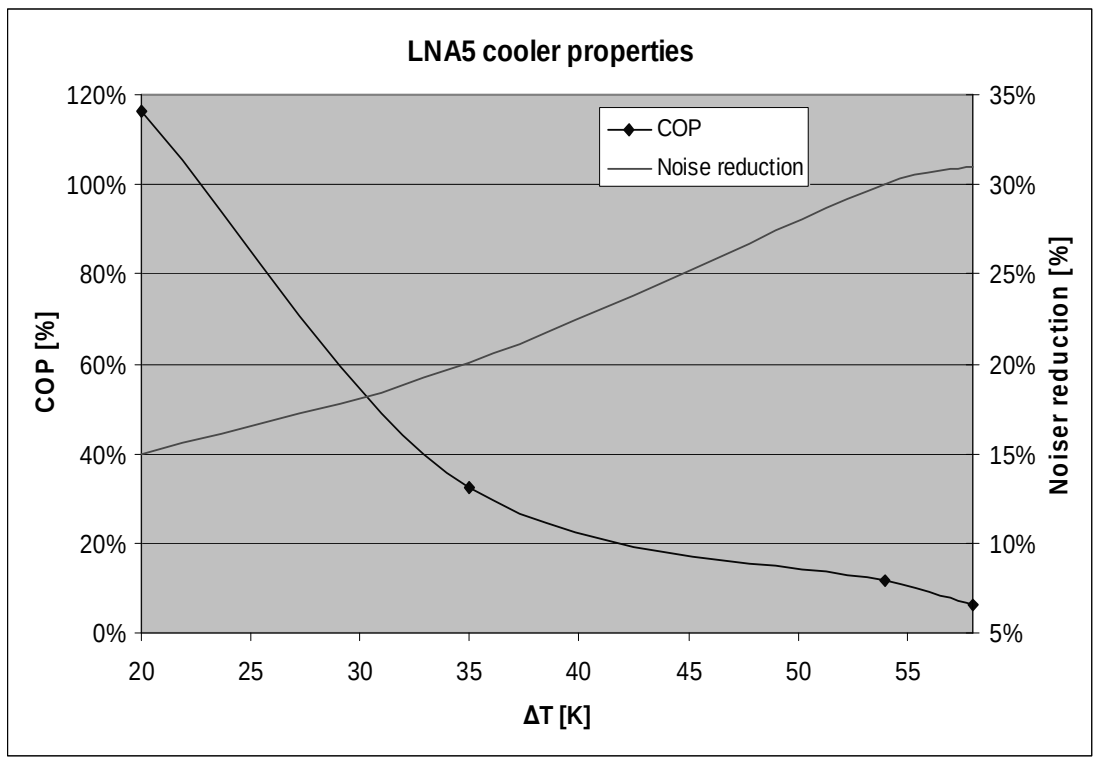

**Figure 8: Peltier COP and noise reduction vs. ΔT**

Many people think that LNA cooling can not be implemented in SKA because the power consumption is too high. This statement is partly true because cooling to the minimum temperature that this Peltier element can achieve would cost around 0.8W which is 100 MW for the entire SKA telescope because it contains 1 million antenna tiles with 128 LNAs each.

In Figure 7, the noise temperature is plotted versus the Peltier power in watts. This graph shows that the noise drops rapidly at a very low Peltier power. With 30mW of Peltier power, the noise temperature drops with more than 15%. This means that the total effective area of SKA can be reduced by 15% if the local cooling is applied on the first stage LNAs with just 3.8MW of Peltier power, which will only be a minor part of the total required power for SKA.

The noise can even be reduced by 20% if 0.15W of power is applied to the Peltier element.

Figure 8 displays the coefficient of performance (COP) of the Peltier element, this is simply pumped heat power divided by the electric power that is used by the Peltier element. Also the percentage of noise reduction (noise temperature) has been plotted in this graph. As an x-axis, the temperature difference has been used in stead of the Peltier power, because this better describes the actual situation; the noise reduction is also dependent on temperature.

With a very low input power of 35mW, a COP of 120% and a noise reduction of 15% have been realized.

## 8. CONCLUSION

The sensitivity of the SKA telescope can be improved by reducing the noise figure of the total system, $A_{eff}/T_{sys}$ can stay constant while reducing the effective area of the telescope.

Thermal simulations predict that the minimum temperature of the LNA die can be cooled down to 250K with a noise reduction of 15%.

A test setup has been designed for temperature and noise measurements on the LNA cooler, consisting of an aluminium cooling block with a Peltier cooler, an LNA chip and a PCB with impedance matching circuitry. The lowest temperature that has been measured on the cold side of the Peltier element was 240K though if the power of the LNA is turned on, the temperature rises to 250K. When the LNA was cooled down to 250K, a noise decrement of 30% has been measured in stead of the simulated 15% at 250K. This high noise decrement is caused by the shot noise which is generated by the gate leakage current in the LNA. The gate leakage current is also strongly dependent on the temperature and gives a 2$^{nd}$ order slope to the temperature dependency of the noise figure.

Power consumption of the total system of SKA is an important issue in the SKA project since the power consumption of every single element must be multiplied by 128.000.000 because the design concept contains 1 million antenna tiles with 128 antenna elements each. A power consumption of 0.5W for the Peltier element would therefore be too much for realistic power consumption. Though the lowest temperature (and noise) can be realized if 0.5W of power is applied to the thermoelectric cooler, the cooler cools down to 275K if only 30mW is applied. With this 30mW of Peltier power consumption, 15% noise temperature reduction was measured. With 150mW of Peltier power, 20% of the noise can be reduced.

With the measured reduction of noise with only small power consumption, the effective area of SKA can also be reduced with the same power consumption or even less; local cooling in vacuum can deliver a significant cost reduction this way.